\documentclass[PP,finalmode]{AEA}

\usepackage{mathptmx}

\usepackage{graphicx}

\usepackage{natbib}
\usepackage{booktabs}
\usepackage{tabularx}

\draftSpacing{1.5}
\usepackage{caption}
\usepackage{subcaption}
\begin{document}

\title{Measuring Changes in Disparity Gaps: An Application to Health Insurance}
\shortTitle{Measuring Changes in Disparity Gaps}
\author{Paul Goldsmith-Pinkham, Karen Jiang, Zirui Song, and Jacob Wallace\thanks{Goldsmith-Pinkham: Yale School of Management, 165 Whitney Avenue, New Haven, CT 06511, e-mail: paul.goldsmith-pinkham@yale.edu. 
Jiang: Yale University, PO Box 208421, New Haven, CT 06520. e-mail: karen.jiang@yale.edu. Song: Harvard Medical School, 180 Longwood Avenue, Boston, MA 02115. e-mail: Song@hcp.med.harvard.edu. Wallace: Yale School of Public Health, 60 College Street, New Haven, CT 06520. e-mail: jacob.wallace@yale.edu.}}\pubMonth{May}
\pubYear{2022}
\pubVolume{112}
\maketitle

It is now widely recognized that achieving health equity for members of racial and ethnic minority groups requires isolating and intervening on the manipulable factors that underlie health inequities \citep{bailey2017structural}. The key empirical challenge---identifying the sources of differences across groups (typically means)---echoes work in other contexts outside of health---e.g., the gender wage gap \citep{blau2017gender}; or the racial wealth gap \citep{derenoncourt2021racial}. One approach in this literature is the Kitagawa-Oaxaca-Blinder  approach \citep{kitagawa1955components,oaxaca1973male, blinder1973wage}, which measures how much mean differences across groups in relevant covariates can explain the gaps. 

While the Kitagawa-Oaxaca-Blinder (KOB) approach has been used to decompose mean differences between groups in observational settings  \citep{fortin2011decomposition} or as a tool for estimating treatment effects \citep{kline2011oaxaca,sloczynski2015oaxaca}, in this paper we show that KOB-style decompositions can also be used to decompose sources of treatment effect heterogeneity in policies that reduce disparities, with a focus on identifying sources of treatment effect heterogeneity between groups. Our approach is closely related to work on decomposing sources of treatment effect variation \citep{heckman1997making,djebbari2008heterogeneous,ding2019decomposing,currie2020caused}, but also to a more recent literature on how to estimate heterogeneity in causal effects \citep[e.g., ][]{athey2016recursive,wager2018estimation,denteh2021ml}. 

We first show that the reduction in disparities between groups caused by a treatment can be written as the difference in conditional average treatment effects (CATEs) for each group. Then, we highlight that these CATEs can be decomposed into CATEs driven by other observables (e.g., the ``endowment'' difference) and unexplained differences between groups. We argue that reporting the share driven by each component can be an important summary statistic for researchers interested in understanding group differences, since it forces researchers to focus on characteristics that are causally manipulable, rather than features like gender or race. Finally, we apply this approach to study the impact of Medicare on racial and ethnic disparities in healthcare access and outcomes.

\section{Heterogeneous Treatment Effects and Disparity Gaps}

Consider a population of individuals with randomly assigned treatment $D_{i} \in \{0,1\}$, a discrete multivalued set of controls $X_{i}$ and a set of potential outcomes $Y_{i}(1)$ and $Y_{i}(0)$, with $Y_{i} = D_{i}Y_{i}(1) + (1-D_{i})Y_{i}(0)$. We define a binary group indicator $W_{i}$ such as race or gender.  We assume strong ignorability:

\begin{equation}
    (Y_{i}(0), Y_{i}(1), X_{i}, W_{i} ) \perp D_{i}
\end{equation}

Researchers are often estimate the effect of the treatment $D_{i}$ for each group $W_{i}$ and compare the treatment effect heterogeneity for each group.  We first show that the difference in treatment effects across group $W$ can be written as the difference in the \emph{outcome gap} across $W_{i}$ before and after the treatment.  In other words, the difference in treatment effects across groups is exactly the change in disparity gaps. 

Let $\mu_{d}(w) = E(Y_{i} | D_{i} = d, W_{i} = w)$, and $\mu_{d}(w, x)  = E(Y_{i} | D_{i} = d, W_{i} = w, X_{i} = x)$. Then, $\tau = \mu_{1} - \mu_{0}$, $\tau(w) = \mu_{1}(w) - \mu_{0}(w)$ and $\tau(w,x) = \mu_{1}(w, x) - \mu_{0}(w, x)$. Here, $\tau$ identifies the ATE and $\tau(w)$ and $\tau(w,x)$ each identify different conditional average treatment effects (CATE), where we assume overlap is satisfied. Finally, let $\gamma_{d} = \mu_{d}(1) - \mu_{d}(0)$ be the disparity gap between the two groups in either the control or treatment case.

Then, we note that the change in the disparity gap is exactly the difference in treatment effects:
\begin{align}
    \delta &= \gamma_{1} - \gamma_{0}=  \tau(1) - \tau(0) .
\end{align}

Rewriting group differences in this way is powerful. Since most researchers consider $W_{i}$ which are not causally manipulable (e.g. race or gender) under the ``no causation without manipulation'' maxim of \cite{holland1986statistics}, $\gamma_{d}$ is not a well-defined causal object. But, $\tau_{w}$ is a meaningful causal object with clear interpretation. When comparing $\tau_{1}$ and $\tau_{0}$, the change in the gap between the groups can be re-interpreted as the difference in treatment effects, such that decomposing how the heterogeneity affects disparity gaps can be informative about the source of the disparities themselves. 

Next, we can rewrite this change in the overall gap as a combination of two pieces, analogous to a Kitagawa-Oaxaca-Blinder (KOB) decomposition: a weighted combination of different heterogeneous effects within characteristics, plus an adjustment given a difference in composition of characteristics $x$. Let $\pi_{w}(x) = Pr(X_{i} = x | W_{i} = w)$ and consider the following decomposition:
\begin{align}
    \delta &= \sum_{x} \tau(1,x)\pi_{1}(x)- \tau(0,x)\pi_{0}(x)  \\
    &=    \sum_{x}\delta(x)\pi_{1}(x) \label{decomp2}\\
    & \qquad +  \sum_{x}\tau(0,x)(\pi_{1}(x)-\pi_{0}(x)). 
\end{align}

Note that these are two types of disparity gaps: one is a difference in outcomes \emph{given} a characteristic, and the other is a gap due to differences across groups in observable characteristics. Since $\delta(x)$ can be defined as either a  difference in treatment effects, or a change in disparity gaps following a treatment, it is possible to interpret the difference in treatment effects as a change in disparities within characteristics (weighted by one group's distribution of characteristics) and the pre-existing difference in characteristics that affect treatment (weighted by one group's treatment effect). An important caveat is that alternative weightings are feasible based on whether the analysis uses $w = 1$ or $w = 0$ as the reference group (as with KOB decompositions). Researchers are encouraged to confirm that the implication of the results below do not shift meaningfully depending on the choice of reference.

A natural measure for this heterogeneity is $\kappa = \sum_{x} \delta(x)\pi_{1}(x)  \big/ \delta$. $\kappa$ can be larger than 1 or less than 0.  Two extreme cases are worth noting: 
\begin{enumerate}
    \item $\tilde{\delta} = 0$. Then, all effects are driven by cross-characteristics composition: $\delta = \sum_{x} \tau(x) (\pi_{1}(x) - \pi_{0}(x)) $, and $\kappa = 0$.
    \item $\pi_{1}(x) - \pi_{0}(x) = 0$ for all $x$. Then, $\delta = \sum_{x} \delta(x) Pr(X_{i} = x)$, and all effects are driven by within-characteristic heterogeneity differences. Then, $\kappa = 1$.
\end{enumerate}

There are two important things to note about use of this decomposition. First, if a policymaker considers targeting the treatment, a large $\kappa$ suggests that benefits accrue \emph{within}-characteristics. As a result, policy makers will achieve reductions in disparity gaps by ensuring take-up of the treatment within a given place where $\delta(x)$ is large. Of course, focusing on disparities may ignore cases where both $\tau_{1}(x)$ and $\tau_{0}(x)$ are large, but the difference $\delta(x)$ is small. In contrast, if $\kappa$ is small, disparities can be reduced by targeting the treatment to where the covariance between $\tau(x)$ and $\pi_{1}(x) - \pi_{0}(x)$ is positive.

Second, consider the setting when $\kappa =0$. In this setting, is it possible to say that differences in $x$ cause the initial disparity gap? In other words, if we causally manipulated $X$ such that the distribution of $X$ was identical across $W$, how would this change $\gamma_{0}$? Without additional assumptions, this is not knowable. If individuals sort by $X$, then our estimates of $E(Y | X_{i} = x)$ do not identify the counterfactual average for all individuals. 

\section{Data and Empirical Framework}

We use this approach to study the impact of eligibility for Medicare---(nearly) universal public health insurance that most Americans qualify for beginning at age 65---on racial and ethnic disparities in healthcare access and outcomes.

The data used in the study come from the Behavioral Risk Factor Surveillance System (BRFSS) for the period 2010-2018, a health-related telephone survey that collects demographics and health-related outcomes from individuals in all 50 states \citep{brfss}. See the Online Appendix for additional details on the BRFSS. Importantly, the data include age in years and state identifiers, allowing us to estimate the local average treatment effect of eligibility for Medicare at age 65 in each region of the country using a regression discontinuity design \citep{goldsmith2020medicare,wallace2021changes}.\footnote{However, in principle these decomposition methods could be used in the context of program evaluations using a range of research designs (e.g., randomized controlled trials, difference-in-differences, etc.).}

In our application, we limit the data to respondents aged 51-79 years and estimate regression discontinuity analyses (see the Online Appendix for additional details on the estimation) both at the national and region levels, and separately by race/ethnicity, using equations of the form:
\begin{align}
\begin{split}
\label{eq:rd_locations}
y_{i,j,l}(\text{age}) = &\tau \times 1(\text{age}>65) \\
&+ f\left(\text{age}\right)\times 1(\text{age}\leq 65) \\
&+ g\left(\text{age}\right)\times 1(\text{age}>65) \\
&+ \epsilon_{i,t,l}(\text{age}).
\end{split}
\end{align}
where $y_{i,j,l}(\text{age})$ is an outcome for individuals $i$ of type $j$ (i.e., self-identified race/ethnicity) in location $l$ of a given age. The functions $f\left(\text{age}\right)$ and $g\left(\text{age}\right)$ are the age profile of $y_{i,j,l}$ for those below and above age 65, respectively, and adjust for the effect of age on our outcomes. The two outcomes to illustrate our approach are the share of survey respondents that report being unable to see a physician in the past year due to cost and the share of respondents with any source of health insurance coverage. We report additional outcomes in the Online Appendix.

Using this framework, we can estimate $\tau$, the treatment effect of Medicare ($D_{i}$), for each outcome. Likewise, for each outcome we can also estimate $\tau(w,x)$ and $\tau(w)$ by re-estimating the RD regression within those groups. This provides the necessary pieces to estimate $\delta$ and $\kappa$.

\section{Results}

\begin{table}[t]
\footnotesize
\caption{Changes in Racial and Ethnic Gaps and the Effects of Medicare Eligibility at Age 65}
\label{table}
 \begin{tabular}{lrrrrrrr}
& \multicolumn{2}{c}{Share of Change in} & \multicolumn{3}{c}{   } & \multicolumn{2}{c}{   }  \\ 
& \multicolumn{2}{c}{ Gap Explained ($\kappa$) } & \multicolumn{3}{c}{Estimated Effect ($\tau$)} & \multicolumn{2}{c}{Change in Gap  ($\delta$)}  \\ 
 \cmidrule(lr){2-3}  \cmidrule(lr){4-6}  \cmidrule(lr){7-8}
Share unable to see physician & \multicolumn{1}{c}{Black}  & \multicolumn{1}{c}{Hispanic} & \multicolumn{1}{c}{White} & \multicolumn{1}{c}{Black}  & \multicolumn{1}{c}{Hispanic}  & \multicolumn{1}{c}{Black}  & \multicolumn{1}{c}{Hispanic}  \\
in the past year due to cost  & \multicolumn{1}{c}{Americans} &  \multicolumn{1}{c}{Americans} &  \multicolumn{1}{c}{Americans} &   \multicolumn{1}{c}{Americans} &  \multicolumn{1}{c}{Americans} & \multicolumn{1}{c}{Americans} &  \multicolumn{1}{c}{Americans}\\
\cmidrule(lr){2-3} \cmidrule(lr){4-6} \cmidrule(lr){7-8}
& \multicolumn{1}{c}{(1)} & \multicolumn{1}{c}{(2)} & \multicolumn{1}{c}{(3)} & \multicolumn{1}{c}{(4)} & \multicolumn{1}{c}{(5)} & \multicolumn{1}{c}{(6)} & \multicolumn{1}{c}{(7)} \\
\cmidrule(lr){2-3} \cmidrule(lr){4-6} \cmidrule(lr){7-8}
\cmidrule(lr){1-1}
\hspace{20pt} \textit{Overall}  & & &  -0.03  &   -0.05   &     -0.07 & -0.02 & -0.04\\
& & & (0.003) & (0.01) & (0.01) & (0.01) & (0.015) \\
\cmidrule(lr){1-3}
\hspace{10pt} Breakdown by Region & 0.97 & 1.0 &&&&&\\
\cmidrule(lr){1-3}
\hspace{20pt}  \textit{Non-South}  & & & -0.03  &   -0.04    &    -0.03  &-0.01 & -0.01\\
& & & (0.004) & ( 0.01) & (0.01) & (0.01) & (0.02) \\
\hspace{20pt} \textit{South}  & & & -0.03   &  -0.05    &    -0.12  &-0.02&  -0.09\\
& & & (0.005) & (0.02) & (0.03) & (0.02) & (0.04) \\
\cmidrule(lr){1-3}
Share Insured &&&&&&& \\
\cmidrule(lr){1-1}
\hspace{20pt} \textit{Overall}  & & &  0.063  &   0.092   &     0.15 & 0.029 & 0.083\\
& & & (0.003) & (0.01) & (0.01) & (0.01) & (0.015) \\
\cmidrule(lr){1-3}
\hspace{10pt} Breakdown by Region & 0.85 &  0.99 &&&&& \\
\cmidrule(lr){1-3}
\hspace{20pt}  \textit{Non-South}  & & & 0.056  &   0.084    &    0.1  &0.028 &  0.047\\
& & & (0.003) & ( 0.01) & ( 0.02) & (0.02) & (0.02) \\
\hspace{20pt} \textit{South}  & & & 0.076    &  0.097     &    0.21  &0.021&  0.14\\
& & & (0.005) & (0.01) & (0.03) & (0.02) & (0.03) \\
\bottomrule
\end{tabular}
\begin{tablenotes}[Source]
\footnotesize
Authors analysis of the Behavioral Risk Factor Surveillance System (BRFSS), 2010-2018. This table reports estimated effects from the Medicare RD regression, for White, Black, and Hispanic Americans. Columns 1 and 2 explain the share of the change in the gap explained by heterogeneity within covariates (see text for details). Columns 3, 4 and 5 report estimates of the RD regression (equation \ref{eq:rd_locations}) by racial and ethnic group. Columns 6 and 7 report the difference between columns 4 and 3, and 5 and 3, respectively. These can be interpreted as the discontinuity in the disparity at age 65.
\end{tablenotes}
\end{table}
 
An advantage of the RD approach used in this application is that it allows us to easily visualize the treatment effects $\gamma$ graphically, as well as the extent to which the treatment reduces racial and ethnic disparities at age 65. In Figure 1 of the Online Appendix, we report plots of the RD estimates separately by race/ethnicity and for states grouped into three groups: ``All States'',  ``Southern States'' and ``Other States.'' We see clear evidence of pre-Medicare disparities, such as in Panel A, where prior to being eligible for Medicare at age 65, there are visible racial and ethnic differences in the share of respondents that report being unable to see a physician in the past year due to cost, with the rates being highest for Hispanic respondents (27.3\%) in Southern States and lowest for non-Hispanic, White respondents (7.7\%) in Other States.

At the national level, we see a discontinuous reduction at age 65 in the Hispanic-White gap in the share of respondents unable to see a physician due to cost (Table \ref{table}). As we already noted above, the decline in the Hispanic-White gap (4.0pp) can be expressed as the difference in the CATEs for Hispanic respondents (7.0pp) and non-Hispanic White respondents (3.0pp). Relative to the pre-65 gap of 17.9 p.p., this represented a 25\% reduction in the disparity.

We then perform a KOB-style decomposition using our region-level estimates to assess how much of the national-level reduction in the Hispanic-White gap is driven by a reduction in the gap within each region (i.e., ``Southern States'' vs. ``Other States'') at age 65 and how much is driven by treatment effect heterogeneity across regions and differences in where racial and ethnic groups reside (e.g., Black Americans are concentrated geographically in the South). In principle, this decomposition could be performed at a more granular level (e.g., states or counties) but for illustrative purposes we focus on two regions in this application.

For each of the outcomes we examined, the share of the national disparity reduction driven by within-region reductions in racial and ethnic disparities, $\kappa$, was large. For example, these estimates ranged from 0.79 for the Black-White gap in the share of the population without a usual source of care to approximately 1.0 for the Hispanic-White gap in all four outcomes. The, perhaps surprising, implication of this is that the reductions in disparities at the national-level are not the result of Medicare having larger effects in a particular region (e.g., the South) where members of racial and ethnic minority groups are more likely to reside. Instead, it is driven by large reductions in the Black-White and Hispanic-White gaps \textit{within regions} at age 65.

\section{Conclusion}

We show that differences in treatment effects across groups can be interpreted as changes in disparity gaps. Moreover, we show the KOB approach can be extended to decompose differences in disparity gaps into differences in the observable characteristics correlated with treatment effect heterogeneity (e.g., region in our example) and treatment effect heterogeneity between groups (e.g., non-Hispanic White vs. Hispanic respondents) within a region. 

This method may be particularly useful for researchers focused on understanding how treatment effect heterogeneity leads some policies to be disparity-reducing (e.g., Medicare eligibility) and interested in decomposing the sources of treatment effect variation between groups (e.g., by race, gender, and their intersections). If observable characteristics only explain a limited amount of the heterogeneity in the effect of a policy, that suggests there may exist important (currently unobserved) characteristics that differ by group. Researchers should attempt to identify these to better understand the disparate impacts of policies between groups. Researchers may want to split by more characteristics, potentially introducing additional noise. Further work should explore the use of machine techniques such as causal forests to implement this \citep{wager2018estimation}.

\singlespacing
\bibliographystyle{aea}
\bibliography{BibFile}

\end{document}